\documentclass[12pt]{article}
\usepackage{amssymb,amsmath,cite,epsfig}
\usepackage[top=2.5cm, bottom=2.5cm, left=1.5cm, right=1.5cm]{geometry}
\usepackage{indentfirst}
\usepackage{subfig}
\begin{document}
\begin{titlepage}
\begin{flushright}
\end{flushright}
\vspace*{3cm}
\begin{center}
{\Large \textsf{\textbf{On Modified First Law of Black hole Thermodynamics in The Non-Commutative Gauge Theory}}}
\end{center}
\par \vskip 5mm
\begin{center}
{\large \textsf{Abdellah Touati$^{a,b}$, Slimane Zaim$^a$}}\\\vskip
5mm
$^a$D\'{e}partement de Physique, Facult\'{e} des Sciences de la Mati\`{e}re,\\
Universit\'{e} de Batna $1$, Algeria. \\
$^b$Laboratoire de Physique des Rayonnements et de leurs Intéractions avec la Matière, Département de Physique, Faculté des Sciences de la Matière Université de Batna-1, Batna 05000, Algeria\\\vskip
5mm
Email: touati.abph$@$gmail.com, zaim$69$slimane$@$yahoo.com
\end{center}
\par \vskip 2mm
\begin{center} {\large \textsf{\textbf{Abstract}}}\end{center}
\begin{quote}
\setlength{\parindent}{3ex}{
In this paper, we investigated the thermodynamic properties of Schwarzschild black hole (SBH) in the non-commutative (NC) gauge theory of gravity. According to our previous work, we modify the first law of the black hole (BH) thermodynamics by the physical quantity (NC potential) $\mathcal{A}$ which is the conjugate to the NC parameter $\Theta$, which leads to this expression $d\hat{M}=\hat{T}d\hat{S}+\mathcal{A}d\Theta$. Our result shows that the NC SBH has a phase transition, and the non-commutativity affected this transition. And the NC potential $\mathcal{A}$ is effective only in the final stage of the BH evaporation, where it increases the Gibbs free energy at this stage, and the NC parameter in this study can represent the tension of the spacetime. Then the study of the pressure of the SBH in the modified first law of the BH thermodynamic shows a second-order phase transition, and the critical value of the thermodynamical variables are related to each value of $\Theta$, and that leads to this parameter to play the same role as a thermodynamical variable.}
\end{quote} \vspace*{2cm}
\begin{quote}
%\textbf{\sc Pacs numbers}: 11.10.Nx, 11.15.-q, 03.65.Pm
\end{quote}
\end{titlepage}

\section{Introduction}

\setlength{\parindent}{3ex}
Quantum gravity is one of the biggest problems of modern physics and one of the large research areas. At the end of the last century, several theories have emerged, to solve this problem. Among the most promising theories are String theory \cite{ST1,ST2}, Loop quantum gravity \cite{LQG1,LQG2} as well as Super-gravity \cite{SG1,SG2,SG3,SG4}. Despite the impressive application and theoretical results provided by these theories, none of these provide a complete theory of quantum gravity and until this day, none of these have been confirmed by experiment. In addition to previous theories, another way of quantifying gravity has been proposed, known as the non-commutative gauge theory of gravity \cite{cham1}. As in quantum mechanics, the dynamical quantities are non-commutative which is the commutation relations between position and momentum, and on the other hand, the gravity in general relativity told us the dynamical quantities are the spacetime. So is natural that the search for the quantum spacetime is obtained by enforcing commutation relations between position coordinates themselves.
\begin{equation}\label{eq:1}
	\left[\hat{x}^\mu,\hat{x}^\nu\right]=i\Theta^{\mu\nu},
\end{equation}
where $\theta^{\mu\nu}$ is an anti-symmetric real matrix.
Then we have the Heisenberg uncertainty relations between position coordinates themselves, which is analog to the ordinary one between position and momentum coordinates. The concept of quantum gravity in this approach is resumed in an idea that states the quantifying of space-time leads to quantifying gravity. Non-commutativity is mainly motivated by string theory \cite{seiberg1} and its natural consequence \cite{berger}. The ordinary product between functions in non-commutative space is replaced by the $\star$-product, where the $\star$-product between
two arbitrary functions $f\left(x\right)$ and $g\left(x\right)$ is given by:
\begin{equation}\label{eq:2}
	f\left(x\right)\star g\left(x\right)=f\left(x\right)\cdot g\left(x\right)+\left.\exp\left(\frac{i}{2}\,\Theta^{\mu\nu}\,\partial_i\,\acute{\partial}_j\right)f\left(x\right) g\left(\acute{x}\right)\right\vert_{x=\acute{x}}.
\end{equation}
Here one uses a gauge field theory with the star product and Seiberg-Witten (SW) maps \cite{seiberg1}. However, this theory of NC gravity is not the final theory of quantum gravity as the one we selected above. The quantum gravity effect due to the non-commutativity of the spacetime is appear in a strong gravitational field, which means that the best application of this theory is the study of black holes. The black hole thermodynamics play an important role in modern physics, and provide a real connection between gravity and quantum mechanics \cite{hawking1,hawking2,hawking3}. Recently there has been a lot of interest in modified thermodynamics proprieties of the black hole by introducing the non-commutativity \cite{piero1,lopez1,nozari1,nozari2,myung1,chai2,mukhe1,zaim1,linares,abdellah2}.

In the present study, we propose a deformed SBH solution in the NC gauge theory of gravity. We apply the Bekenstein-Hawking method to compute the thermodynamical properties. Then we modified the first law of the BH thermodynamics by the non-commutativity of the space-time. Our result shows that the NC potential $\mathcal{A}$ is active only in the final stage of the SBH evaporation. And the NC parameter $\Theta$ plays the same role as a thermodynamical variable in the presence of the pressure.

This paper is organized as follows. In Sect. 2 we present the NC gauge gravity metric for SBH by using S-W maps, following the approach of Ref. \cite{cham1}, and we calculate the corrected event horizon radius up to second order in NC parameter $\Theta$. In Sect. 3 we calculate the NC ADM mass, Hawking temperature, and entropy of the deformed SBH in the NC spacetime where we stop the correction up to the second-order in $\Theta$. In section 4 we investigated the modification in the first law of the BH thermodynamics by the NC potential and NC charge (parameter $\Theta$) and their influence on the phase transition.

\section{Non-commutative corrections for Schwarzschild black hole}

In our previous works \cite{abdellah1,abdellah2}, we used the tetrad formalism and both the star $\ast-$product and the SW map to construct a non-commutative gauge theory for a static metric with spherical symmetric. One can use a perturbation form for the SW map to describe the deformed tetrad fields $\hat{e}^{a}_{\mu}(x,\Theta)$ as a development in the power of $\Theta$ up to the second-order, which can be obtained by following the same approach in Ref. \cite{cham1}: 
\begin{align}
\hat{e}_\mu^a=\,&e_\mu^a-\frac{i}{4}\Theta^{\nu\rho}[\omega^{ac}_{\nu}\partial_{\rho}e^{d}_{\mu}+(\partial_{\rho}\omega^{ac}_{\mu}+R^{ac}_{\rho\mu})e^{d}_{\nu}]\eta_{cd}\notag\\
&+\frac{1}{32}\Theta^{\nu\rho}\Theta^{\lambda\tau}\left[2\{R_{\tau\nu},R_{\mu\rho}\}^{ab}e^{c}_{\lambda}-\omega^{ab}_{\lambda}(D_{\rho}R_{\tau\nu}^{cd}+\partial_{\rho}R_{\tau\nu}^{cd})e^{m}_{\nu}\eta_{dm}\right.\notag\\
&\left.-\{\omega_{\nu},(D_{\rho}R_{\tau\nu}+\partial_{\rho}R_{\tau\nu})\}^{ab}e^{c}_{\lambda}-\partial_{\tau}\{\omega_{\nu},(\partial_{\rho}\omega_{\mu}+R_{\rho\mu})\}^{ab}e^{c}_{\lambda}\right.\notag\\ 
&\left.-\omega^{ab}_{\lambda}\left(\omega^{cd}_{\nu}\partial_{\rho}e^{m}_{\mu}+\left(\partial_{\rho}\omega_{\mu}^{cd}+R_{\rho\mu}^{cd}\right)e^{m}_{\nu}\right)\eta_{dm}+2\partial_{\nu}\omega_{\lambda}^{ab}\partial_{\rho}\partial_{\tau}e^{c}_{\lambda}\right.\notag\\
&\left.-2\partial_{\rho}\left(\partial_{\tau}\omega_{\mu}^{ab}+R_{\tau\mu}^{ab}\right)\partial_{\nu}e^{c}_{\lambda}-\{\omega_{\nu},(\partial_{\rho}\omega_{\lambda}+R_{\rho\lambda})\}^{ab}\partial_{\tau}e^{c}_{\mu}\right.\notag\\
&\left.-\left(\partial_{\tau}\omega_{\mu}+R_{\tau\mu}\right)\left(\omega^{cd}_{\nu}\partial_{\rho}e^{m}_{\lambda}+\left((\partial_{\rho}\omega_{\lambda}+R_{\rho\lambda})\right)e^{m}_{\nu}\right)\eta_{dm}\right]\eta_{cb} +\mathcal{O}\left( \Theta^{3}\right),\label{eq:SWM}
\end{align}
where $\hat{e}_{a}^{\mu }$ and $\omega^{ab}_{\mu}$ are the tetrad field and the spin connection (gauge field), and:
\begin{align}
	\{\alpha,\beta\}^{ab}=\left(\alpha^{ac}\beta^{db}+\beta^{ac}\alpha^{db}\right)\eta_{cd},\quad &[\alpha,\beta]^{ab}=\left(\alpha^{ac}\beta^{db}-\beta^{ac}\alpha^{db}\right)\eta_{cd}\\
	D_{\mu}R_{\rho\sigma}^{ab}=\partial_{\mu}R^{ab}_{\rho\sigma}+&\left(\omega_{\mu}^{ac}R^{db}_{\rho\sigma}+\omega_{\mu}^{bc}R^{da}_{\rho\sigma}\right)
\end{align}

where the $\hat{e}_{a}^{\mu }$ is the of the vierbein $\hat{e}_{\mu }^{a}$ defined as:
\begin{equation}
\hat{e}_{\mu }^{b} \hat{e}_{a}^{\mu }=\delta _{a}^{b},\quad \hat{e}_{\mu }^{a} \hat{e}_{a}^{\nu }=\delta _{\mu }^{\nu }\,.
\end{equation}

In the following, we consider a symmetric metric $\hat{g}_{\mu \nu }$, so that:
\begin{equation}\label{eq:22}
\hat{g}_{\mu \nu }=\frac{1}{2}(\hat{e}_{\mu }^{b}\ast \hat{e}_{\nu b}+\hat{e}_{\nu }^{b}\ast \hat{e}_{\mu b})\,.
\end{equation}
To compute the deformed metric $\hat{g}_{\mu \nu }$, we choose the following NC anti-symmetric matrix $\Theta^{\mu\nu}$:
\begin{equation}
	\Theta^{\mu\nu}=\left(\begin{matrix}
		0	& 0 & 0 & 0 \\
		0	& 0 & 0 & \Theta \\
		0	& 0 & 0 & 0 \\
		0	& -\Theta & 0 & 0
	\end{matrix}
	\right), \qquad \mu,\nu=0,1,2,3\label{eqt2.34}
\end{equation}

We follow the same steps outlined in Ref. \cite{abdellah1}, we choose the following general tetrads field:
\begin{align}
\underline{e}_{\mu }^{0}& =\left(\begin{array}{cccc}\left(1-\frac{2 m}{r}\right)^{\frac{1}{2}}, & 0, & 0, & 0\end{array}\right), \\
\underline{e}_{\mu }^{1}& =\left(\begin{array}{cccc}0 & \left(1-\frac{2 m}{r}\right)^{-\frac{1}{2}}sin\theta cos\phi & r cos\theta cos\phi & -r sin\theta sin\phi\end{array}\right), \\
\underline{e}_{\mu }^{2}& =\left(\begin{array}{cccc}0 & \left(1-\frac{2 m}{r}\right)^{-\frac{1}{2}}sin\theta sin\phi & r cos\theta sin\phi & r sin\theta cos\phi\end{array}\right), \\
\underline{e}_{\mu }^{3}& =\left(\begin{array}{cccc}0 & \left(1-\frac{2 m}{r}\right)^{-\frac{1}{2}}cos\theta & -r sin\theta & 0\end{array}\right).
\end{align}
The non-zero components of the NC tetrad fields $\hat{e}^{a}_{\mu}$ are calculated in Ref. \cite{abdellah1}. Then, using the definition \eqref{eq:22}, we obtain the following non-zero components of the non-commutative metric $\hat{g}_{\mu \nu }$ up to second order of $\Theta$, we take the case $\theta=\pi/2$:
\small
\begin{align}
	-\hat{g}_{00}=&\left(1-\frac{2 m}{r}\right)+\left\{\frac{m\left(88m^2+mr\left(-77+15\sqrt{1-\frac{2m}{r}}\right)-8r^2\left(-2+\sqrt{1-\frac{2m}{r}}\right)\right)}{16 r^4(-2m+r)}\right\}\Theta^{2}+\mathcal{O}(\Theta^{4})\label{eq:13}\\
	%%%%%%%%%%%%%%%%%%%%%%%%%
	\hat{g}_{11}=&\left(1-\frac{2 m}{r}\right)^{-1}+\left\{\frac{m\left(12m^2+mr\left(-14+\sqrt{1-\frac{2m}{r}}\right)-r^2\left(5+\sqrt{1-\frac{2m}{r}}\right)\right)}{8r^2(-2m+r)^3}\right\}\Theta^{2}+\mathcal{O}(\Theta^{4})\label{eqt2.49}\\
	%%%%%%%%%%%%%%%%%%%%%%%%
	\hat{g}_{22}=&r^{2}+\left\{\frac{m\left(m\left(10-6\sqrt{1-\frac{2m}{r}}\right)-\frac{8m^2}{r}+r\left(-3+5\sqrt{1-\frac{2m}{r}}\right)\right)}{16(-2m+r)^2}\right\}\Theta^{2}+\mathcal{O}(\Theta^{4})\label{eqt2.50}\\
	%%%%%%%%%%%%%%%%%%%%%%%%
	\hat{g}_{33}=&r^{2}+\left\{\frac{5}{8}-\frac{3}{8}\sqrt{1-\frac{2m}{r}}+\frac{m\left(-17+\frac{5}{\sqrt{1-\frac{2m}{r}}}\right)}{16r}+\frac{m^2\sqrt{1-\frac{2m}{r}}}{(-2m+r)^2}\right\}\Theta^{2}+\mathcal{O}(\Theta^{4})\label{eqt2.51}
\end{align}
\normalsize
It is clear that, for $\Theta\rightarrow0$, we obtain the commutative Schwarzschild solution. For such a black hole, we find that the event horizon in non-commutative space-time is where the NC metric \eqref{eq:13} satisfies the following conditions:
\begin{equation}
\hat{g}_{00}=0\,.
\end{equation}
The solution to this equation gives us the NC event horizon of the Schwarzschild black hole (SBH):
\begin{equation}
	r_{h}^{NC}=r_{h}\left[1+\left(\frac{\Theta}{r_h}\right)\left(\frac{4\sqrt{5}+1}{32\sqrt{5}}\right)+\left(\frac{\Theta}{r_h}\right)^2\left(\frac{10+\sqrt{5}}{128}\right)\right]\label{eq:18}
\end{equation}
where $r_{h}=2m$ is the event horizon in commutative case when $\Theta =0$. It is interesting to note that the NC correction to the event horizon contains the first order in $\Theta$. The effect of non-commutativity is small, which is reasonable to expect since at large distances can be considered neglected.

\section{First law of black hole thermodynamics in the NC gauge theory}

In this section, we use the result of the previous sections to describe the thermodynamic quantities of the NC SBH. As the first step, we compute the ADM mass of the NC SBH as a function of the event horizon $r_h$, which can be derived from the equation $\hat{g}_{00}(r^{NC}_h)=0$, where the solution can be expressed as series:
\begin{equation}
	\hat{M}_{ADM}=M_{ADM}+\Theta M^{(1)}+\Theta^2 M^{(2)}
\end{equation}
where $M^{(1)}$ and $M^{(2)}$ are the NC correction:
\begin{align}	
	M^{(1)}&=\frac{1}{2} \left(\frac{a(-3+128a^2)}{6+256a^2}\right) \label{eq:19}\\
	M^{(2)}&=\left(\frac{1}{2r_h}\right)\left(\frac{3b+a^2(-35+128 b)}{2(3+128a^2)}\right)
\end{align}
and $M_{ADM}=\frac{r_h}{2}$ is the ADM mass of the commutative SBH, and $a=\left(\frac{4\sqrt{5}+1}{32\sqrt{5}}\right)$, $b=\left(\frac{10+\sqrt{5}}{128}\right)$.

The usual Hawking temperature near the horizon of the NC black hole:
\begin{equation}
	\hat{T}_H=\frac{1}{4\pi}\left|\frac{\partial g_{00}}{\partial r}\right|_{r=r_h^{NC}}\approx T_H+\Theta T^{(1)}+\Theta^2 T^{(2)}
\end{equation}
using \eqref{eq:18} with the commutative component in \eqref{eq:13}:
\begin{align}	
	T^{(1)}&=-\frac{1}{4\pi r_h^2}\left(\frac{4\sqrt{5}+1}{16\sqrt{5}}\right) \label{eq:23}\\
	T^{(2)}&=-\left(\frac{1}{4\pi r_h^3}\right)\left(\frac{10+\sqrt{5}}{64}\right)\label{eq:24}
\end{align}
where $T_H=\frac{1}{4\pi r_h}$ is the commutative Hawking temperature. Where the entropy of the NC SBH is computing using the event horizon \eqref{eq:18}:
\begin{equation}
	\hat{S}=\frac{A_h^{NC}}{4}\approx S+\Theta S^{(1)}+\Theta^2 S^{(2)}\label{eq:25}
\end{equation}
with:
\begin{align}	
	S^{(1)}&=\pi r_h \left(\frac{4\sqrt{5}+1}{16\sqrt{5}}\right) \label{eq:26}\\
	S^{(2)}&=\pi\left(\left(\frac{4\sqrt{5}+1}{32\sqrt{5}}\right)^2+\left(\frac{10+\sqrt{5}}{64}\right)\right)\label{eq:27}
\end{align}
where the commutative entropy of SBH $S=\pi r_h^2$.

In our previous work \cite{abdellah2}, we use the usual first law of black hole thermodynamics in the NC spacetime, which is defined as:
\begin{equation}
	d\hat{M}_{ADM}=\hat{T}_Hd\hat{S}\label{eq:28}
\end{equation}
the corresponding integral Bekenstein-Samarr for the ADM mass is:
\begin{equation}
	\frac{1}{2}\hat{M}_{ADM}=\hat{T}_H\hat{S}\label{eq:29}
\end{equation}
Our aim is to separate the commutative terms and the NC terms, to study the NC property of geometry as a thermodynamic variable, for that we use the relations \eqref{eq:22} and \eqref{eq:25}
\begin{align}\label{eq:30}
	\frac{1}{2}\hat{M}_{ADM}&=\left(T_H+\Theta T^{(1)}+\Theta^2 T^{(2)}\right)\left(S+\Theta S^{(1)}+\Theta^2 S^{(2)}\right)\notag\\
	&=T_HS+\Theta\left(T^{(1)}S+T_HS^{(1)}\right)+\Theta^2\left(T_HS^{(2)}+T^{(2)}S+T^{(1)}S^{(1)}\right)+\Theta^3(...)+..\notag\\
	&=T_HS+\Theta\mathcal{A}
\end{align}
where $\mathcal{A}$ describes the NC correction for the first law of black hole thermodynamics and can be seen as a physical quantity conjugate to the NC parameter $\Theta$. From the relation \eqref{eq:30}:
\begin{equation}
	\mathcal{A}=\left(T^{(1)}S+T_HS^{(1)}\right)+\Theta\left(T_HS^{(2)}+T^{(2)}S+T^{(1)}S^{(1)}\right)+\Theta^2(T^{(1)}S^{(2)}+T^{(2)}S^{(1)})+\Theta^3T^{(2)}S^{(2)}\label{eq:31}
\end{equation}
Take now the differential form of the relation \eqref{eq:30}, we get:
\begin{equation}
	d\hat{M}_{ADM}=T_HdS+\mathcal{A}d\Theta\label{eq:32}
\end{equation}
where $T_HdS$ is the commutative case and $\mathcal{A}d\Theta$ has all information about the NC correction. The Gibbs free energy now became:
\begin{equation}
	\hat{G}=\hat{M}_{ADM}-T_H S-\mathcal{A} \Theta\label{eq:33}
\end{equation}
The detailed description of the Gibbs free energy can be found in Ref. \cite{abdellah2}.

\section{Modified first law of the BH thermodynamics}

As we saw earlier in the relation \eqref{eq:32} and \eqref{eq:33}, the NC corrections can be seen as a thermodynamics variable. Like the electric potential in the black hole Reissener-Nordström, as well as every quantum potential used for the corrections in the black hole, all that affects the quantum process of Hawking radiation. However, the deformed geometry also affects this process, which this deformation plays a similar role as potential, so the Hawking radiation is modified by this deformed geometry and that is laid to a modification in the black hole thermodynamics. 

Our aim is to investigate the NC correction as a thermodynamically variable in the first law of BH thermodynamics, which is motivated by our previous work \cite{abdellah2}, where we found that in the phase transition of the NC SBH in presence of the pressure the NC parameter $\Theta$ play the same role as a thermodynamical variable. 

For the SBH in the NC spacetime the first law of the BH thermodynamics \eqref{eq:28} now is modified by the non-commutativity:

\begin{equation}
	d\hat{M}_{ADM}=\hat{T}_Hd\hat{S}+\mathcal{A}d\Theta\label{eq:34}
\end{equation}
where $\Theta$ is the NC parameter and $\mathcal{A}$ his conjugate, which is defined as:
\begin{equation}\label{eq:35}
\mathcal{A}=\left(\frac{\partial\hat{M}_{ADM}}{\partial\Theta}\right)_{\hat{S}}
\end{equation}
use the definition of the ADM mass \eqref{eq:19}, we get:
\begin{equation}\label{eq:36}
	\mathcal{A}=\frac{1}{2} \left(\frac{a(-3+128a^2)}{6+256a^2}\right)+\left(\frac{3b+a^2(-35+128 b)}{4(3+128a^2)}\right)\left(\frac{\Theta}{r_h}\right)
\end{equation}
for numerical application, we get:
\begin{equation}\label{eq:37}
	\mathcal{A}=-c-d\frac{\Theta}{r_h}
\end{equation}
where $c\approx0.003351$ and $d\approx0.006985$, and the NC parameter is positive. As we see the NC potential conjugate to $\Theta$ is similar to the Coulomb potential conjugate to the electric charge.

\begin{figure}[h]
	\centering
	\includegraphics[width=0.4\textwidth]{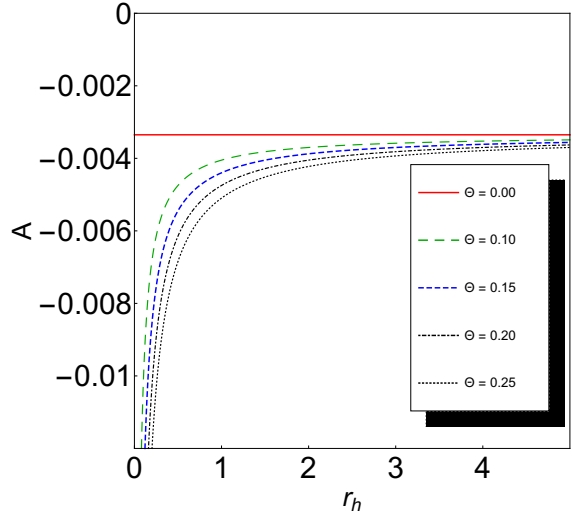}
	\caption{Plot the NC potential $\mathcal{A}$ of the black hole as function of the event horizon $r_h$.}	\label{fig1}
\end{figure}

As we see, for a large BH ($r_h\rightarrow\infty$) the NC potential became constant $\mathcal{A}=-c$, and for the small one, the NC potential became important, which confirmed that the effect of the non-commutativity is local. 

In order to analyze the local and global thermodynamic stability and the phase transition of the SBH in the NC spacetime we investigate the Gibbs free energy:
\begin{equation}
	\hat{G}=\hat{M}_{ADM}-\hat{T}_H \hat{S} \label{eq:38}
\end{equation}
according to the modified first law of BH thermodynamic \eqref{eq:34}:
\begin{equation}
	\hat{G}_{NC}=\hat{M}_{ADM}-\hat{T}_H \hat{S}-\Theta\mathcal{A} \label{eq:39}
\end{equation}

\begin{figure}[h]
	\centering
	\includegraphics[width=0.4\textwidth]{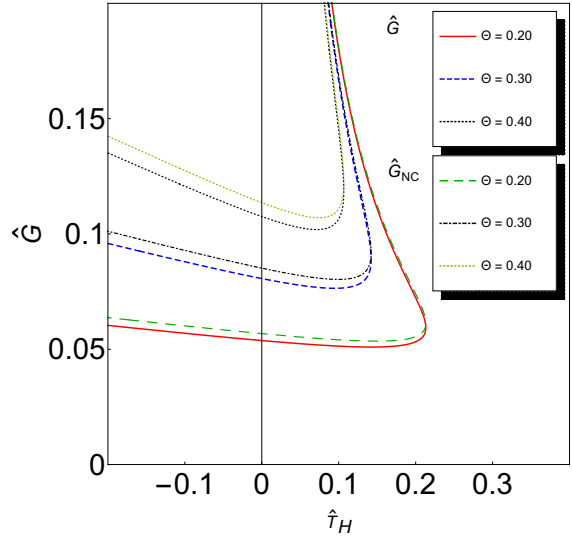}
	\caption{The behaviors of the Gibbs free energy $\hat{G}$ as a function of the NC Hawking temperature $\hat{T}_H$, for the two case \eqref{eq:38} and \eqref{eq:39}.}	\label{fig2}
\end{figure}

In Fig. \ref{fig2} we represent the behavior of the Gibbs free energy for the two cases \eqref{eq:38}, \eqref{eq:39} as a function of the NC Hawking temperature with different values of $\Theta$. As we see the both of Gibbs's free energy decrease rapidly (from the infinity at $\tilde{T}_H \sim	0$) for the low temperature with a small $\Theta$, in this region the NC potential $\mathcal{A}$ is negligible. Then the Gibbs free energy changes its behavior to the increasing in slowly rate at $\tilde{T}_H^{max}\approx0.0426/\Theta$ when the temperature bounces back, with the increase in $\Theta$ this behavior change rapidly. One can see that when the black hole is in the final stage, the effect of the NC potential $\mathcal{A}$ become important and increase the Gibbs free energy of the black hole, and the rate of this growth with the increasing of $\Theta$.
\begin{figure}[h]
	\centering
	\includegraphics[width=0.4\textwidth]{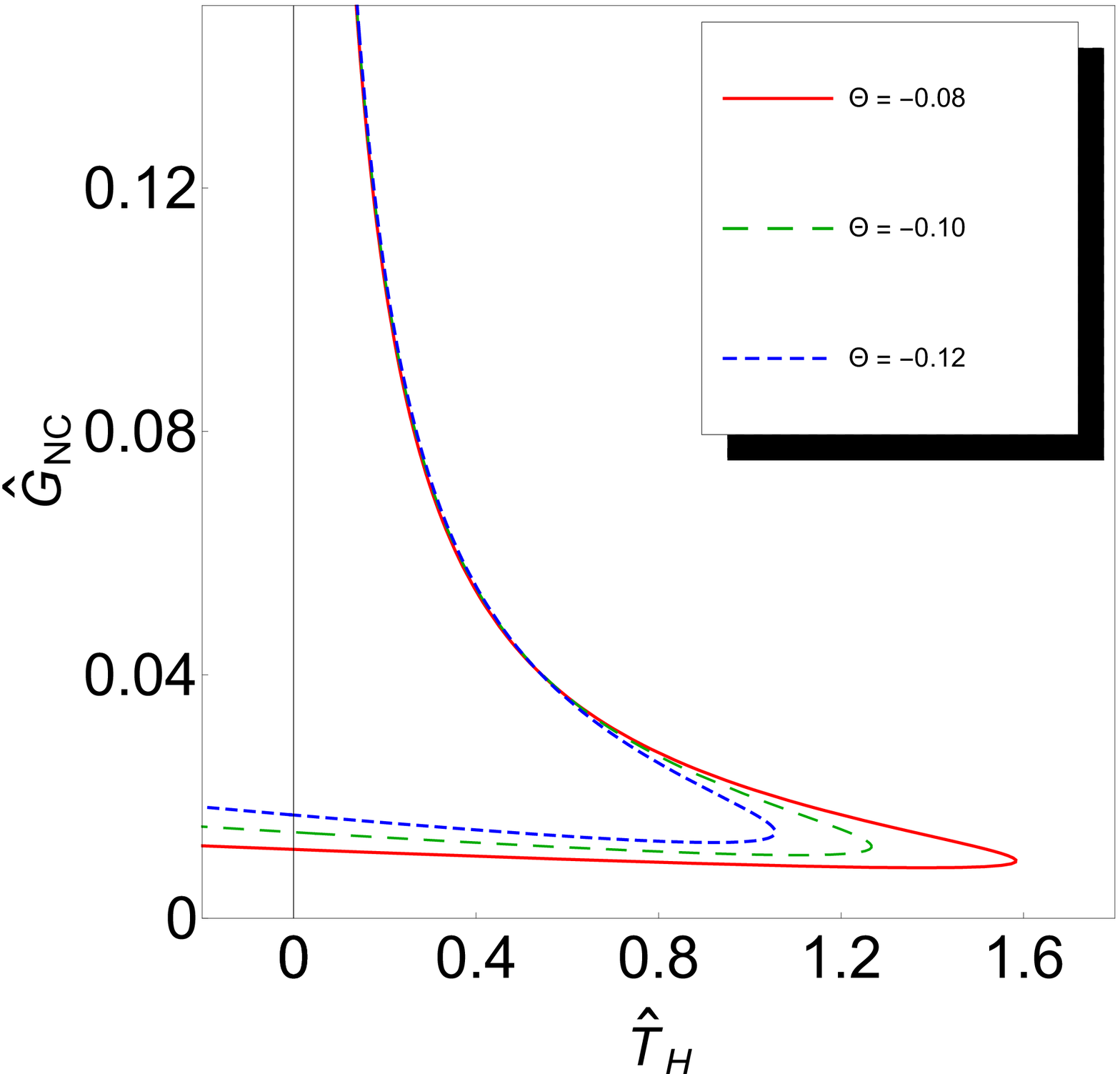}
	\caption{The behaviors of the Gibbs free energy $\hat{G}_{NC}$ as a function of the NC Hawking temperature $\hat{T}_H$ for negative $\Theta$.}	\label{fig3}
\end{figure}

Fig. \ref{fig3} shows the behaviors of the Gibbs free energy as a function of the NC temperature for the negative NC parameter. As we see this behavior is similar to the one obtained by the surface tension in the modified first law of the BH thermodynamics as in Ref. \cite{jawad1,chen1,hansen1}, but in our work, this behavior emerged from the quantum structure of the spacetime on the modified first law of the BH thermodynamic. From this analogy, we can see the NC parameter $\Theta$ as a tension of the spacetime at the quantum scale.  
\subsubsection{Pressure in modified first law of the BH thermodynamics}

Now, will consider the pressure in the NC spacetime and investigate the thermodynamics of the SBH. In the same context as in Ref. \cite{abdellah2}, we study the influence of the pressure on phase transition and the stability condition, but in this work, we use the modified first law of the BH thermodynamics. Let's considered the pressure term in the relation \eqref{eq:34}, so we get:
\begin{equation}\label{eq:40}
	d\hat{M}_{ADM}=\hat{T}_Hd\hat{S}+\mathcal{A}d\Theta-\hat{P}d\hat{V}
\end{equation}
where $\hat{V}$ denote the volume of the NC SBH:
\begin{equation}
	\hat{V}=\frac{4\pi}{3}\left(r^{NC}_h\right)^3\approx V+\Theta V^{(1)}+\Theta^2 V^{(2)} \label{eq:41}
\end{equation}
where $V=\frac{4\pi}{3}r_h^3$ is the commutative volume of the SBH, and:
\begin{align}
	V^{(1)}&=4\pi\left(\frac{4\sqrt{5}+1}{32\sqrt{5}}\right) r_h^2\\
	V^{(2)}&=\pi \left(\left(\frac{4\sqrt{5}+1}{32\sqrt{5}}\right)^2+\left(\frac{10+\sqrt{5}}{128}\right)\right)r_h \label{eq:42}
\end{align}
and $\hat{P}$ his conjugate the pressure of the NC BH:
\begin{align}
	\hat{P}&=-\left(\frac{\partial \hat{M}}{\partial \hat{V}}\right)=-\left(\frac{\partial \hat{M}}{\partial r_h}\right)\left(\frac{\partial \hat{V}}{\partial r_h}\right)^{-1}\notag\\
	&\approx P+\Theta P^{(1)}+\Theta^2 P^{(2)}
\end{align}
where $P=-\frac{1}{8\pi r_h^2}$ is the commutative term of the pressure, with:
\begin{align}
	P^{(1)}&=\frac{a}{4\pi r_h^3}\\
	P^{(2)}&= \left(\frac{(64a^4+3b+16a^2(-1+8b))}{(3+128a^2)}\right)\frac{1}{4\pi r_h^4}
\end{align}

The Gibbs free energy in prisence of the pressure is given by: 
\begin{equation}\label{eq:47}
	\hat{G}=\hat{M}_{ADM}-\hat{T}_H \hat{S}+\hat{P}\hat{V}. 
\end{equation}
According to the modified first law of the BH thermodynamics \eqref{eq:40}, the Gibbs free energy become:
\begin{equation}\label{eq:48}
	\hat{G}=\hat{M}_{ADM}-\hat{T}_H \hat{S}-\Theta\mathcal{A}+\hat{P}\hat{V}.
\end{equation}

\begin{figure}[h]
	\centering
	\includegraphics[width=0.4\textwidth]{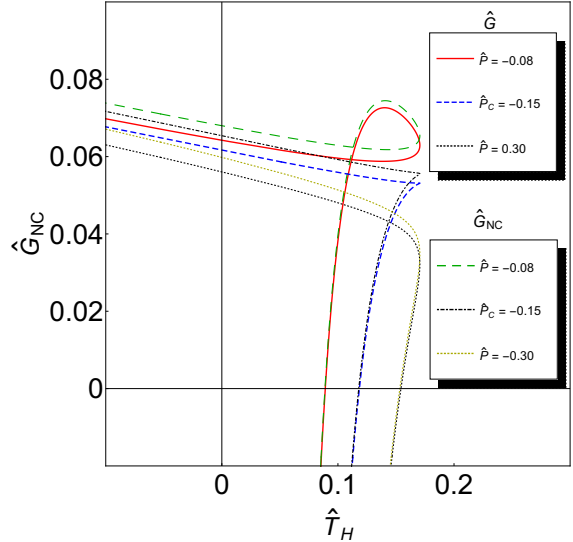}
	\caption{The behaviors of the Gibbs free energy as a function of the NC Hawking temperature $\hat{T}_H$ for the two case \eqref{eq:47} and \eqref{eq:48}, with negative pressure.}	\label{fig4}
\end{figure}
Fig. \ref{fig4} shows the behaviors of the Gibbs free energy in the presence of the pressure as a function of the NC Hawking temperature for the two cases \eqref{eq:47} and \eqref{eq:48}. As we observe, the 'swallow tail' structure appears in the NC spacetime which means two-phase coexist in the first order or can be seen as continuous phase transition as in the literature \cite{chen,hansen,xu,kubiz}.  However, in the NC spacetime, the curves become smoother and conserved the critical point at $\hat{P}_c=-0.15$ with $\Theta=0.25$, where at this critical point the swallowtail structure disappears and an inflection point occurs, which is the second-order phase transition. When $\hat{P}<\hat{P}_c$ these point disappears. All of that is for both of the two cases of the Gibbs free energy \eqref{eq:47} and \eqref{eq:48}. However, in the final stage of the SBH evaporation the NC potential $\mathcal{A}$ increases the Gibbs free energy $\hat{G}_{NC}$. 

\begin{figure}[h]
	\centering
	\includegraphics[width=0.4\textwidth]{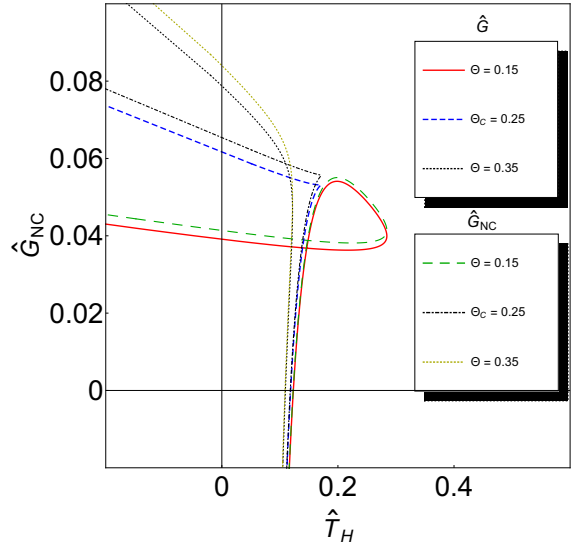}
	\caption{The behaviors of the Gibbs free energy as a function of the NC Hawking temperature $\hat{T}_H$ for the two case \eqref{eq:47} and \eqref{eq:48}.}	\label{fig5}
\end{figure}
Fig. \ref{fig5} shows the behavior of the Gibbs free energy as a function of the NC Hawking temperature $\hat{T}_H$, for different values of $\Theta$. As we see, the swallowtail structure appears for small values of $\Theta$ with smoother curves but conserved the inflection point at the critical point $\Theta_c=0.25$. However, these points disappear when $\Theta>\Theta_c$.
As we see this behavior is similar to the one obtained in Fig. \ref{fig4} but with a difference in the order of the structure of the curve. As we see, for each value of $\Theta$ we have a new critical point of temperature $\hat{T}_C$ and pressure $\hat{P}_C$. In other words, the non-commutativity of the spacetime affects the phase transition of the SBH, we also saw how the NC parameter $\Theta$ he behaves in presence of the pressure, which plays the same role as the thermodynamical variable of the BH. After all, we saw above is interesting to note that the NC potential $\mathcal{A}$ becomes effective only in the final stage of the BH evaporation, and otherwise, it is neglected.

\section{Conclusion}

In this work, we investigated the modified first law of SBH thermodynamics in the NC gauge theory. Our corrections spread the singularity at $r=0$ over a two-dimensional sphere of the radius $r=2m$, where the new event horizon is increasing by the NC correction $r_h^{NC}> r^C_h$, add to that the event horizon contains a first-order in $\Theta$ as all the thermodynamical quantities. 

As a second step, we show that the non-commutativity of the spacetime modified the thermodynamic proprieties of the SBH. Where the ADM mass, the Hawking temperature, and the entropy of the BH are corrected, in order to investigate the influence of the modified first law of BH thermodynamics in the phase transition of the SBH in the NC spacetime. Then we show that the NC parameter can be treated as a thermodynamical variable, by adding a physical quantity $\mathcal{A}$ which is conjugate to the NC parameter $\Theta$. We named this physical quantity "NC potential" because their expression is in a similar form as the Coulomb potential. Our analysis shows that this potential affected only the final stage of the SBH evaporation, where this increases the Gibbs free energy of the BH. 

Finally, the non-commutativity of the spacetime affected the phase transition of the SBH. Where, the NC parameter $\Theta$ in presence of the pressure of the BH plays the same role as the thermodynamical variables, and it can be seen as a tension of the spacetime.

\bibliographystyle{unsrt}
\bibliography{bibliography}
\end{document}